\documentclass[electronic]{vgtc}                 

\ifpdf
  \pdfoutput=1\relax                   
  \usepackage{graphicx}                
  \DeclareGraphicsExtensions{.pdf,.png,.jpg,.jpeg} 
\else
  \usepackage{graphicx}                
  \DeclareGraphicsExtensions{.eps}     
\fi

\graphicspath{{figs/}{figures/}{pictures/}{images/}{./}} 

\usepackage{microtype}                 
\PassOptionsToPackage{warn}{textcomp}  
\usepackage{textcomp}                  
\usepackage{mathptmx}                 
\usepackage{times}                     
         
\usepackage{cite}                      
\usepackage{tabu}                      
\usepackage{booktabs}

\usepackage{amsmath}
\usepackage{amssymb}
\usepackage{graphicx}
\usepackage{algorithmic}
\usepackage{algorithm}
\usepackage{url}
\usepackage{dirtytalk}

\newcommand{\para}[1]       {\vspace{2mm}\noindent{\textbf{#1}}}
\newcommand{\denselist}{\vspace{-5pt} \itemsep -2pt\parsep=-1pt\partopsep -2pt}

\newcommand {\mm}[1] {\ifmmode{#1}\else{\mbox{\(#1\)}}\fi}

\newcommand{\Cech}{\mm{C}}
\newcommand{\Rips}{\mm{R}}

\newcommand{\Ucal}        {\mm{\mathcal U}}
\newcommand{\Dcal}        {\mm{\mathcal D}}

\newcommand{\Rspace}        {\mm{{\mathbb R}}}
\newcommand{\Hgroup}        {\mm{\sf H}}

\onlineid{0}

\vgtccategory{Research}

\vgtcinsertpkg

\title{Visualizing Sensor Network Coverage with Location Uncertainty}

\author{Tim Sodergren\thanks{e-mail: tsodergren@sci.utah.edu}\\ 
        \scriptsize University of Utah 
\and Jessica Hair\thanks{e-mail:jessica.hair@utah.edu}\\ 
     \scriptsize University of Utah 
\and Jeff Phillips\thanks{e-mail:jeffp@cs.utah.edu}\\ 
     \scriptsize University of Utah 
\and Bei Wang\thanks{e-mail:beiwang@sci.utah.edu}\\ 
     \scriptsize University of Utah}

\teaser{
  \centering
  \includegraphics[width=1.0\linewidth]{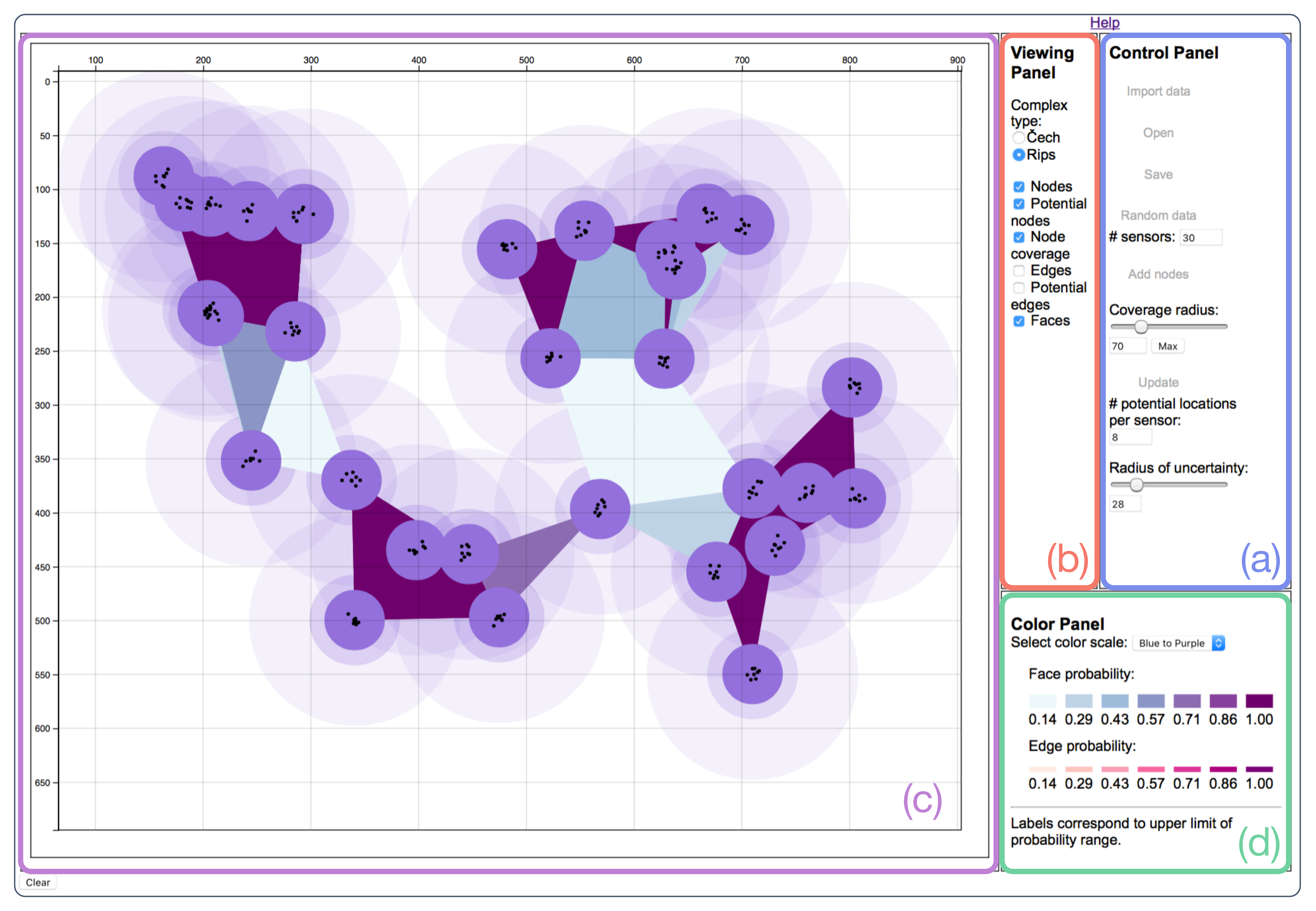}
  \label{fig:teaser}
  \vspace{-8mm}
  \caption{The visual interface in exploring sensor network coverage with location uncertainty: (a) control panel, (b) viewing panel, (c) interaction panel and (d) color panel. In the interaction panel, the coverage of a randomly generated sensor network is visualized by the union of disk-like sensor regions overlaid with a Rips complex representation.}
}

\abstract{
We present an interactive visualization system for exploring the
coverage in sensor networks with uncertain sensor locations. We
consider a simple case of uncertainty where the location of each
sensor is confined to a discrete number of points sampled uniformly
at random from a region with a fixed radius. Employing techniques
from topological data analysis, we model and visualize network
coverage by quantifying the uncertainty defined on its simplicial
complex representations. We demonstrate the capabilities and effectiveness
of our tool via the exploration of randomly distributed
sensor networks.
}

\begin{document}

\maketitle

\section{Introduction}
\label{sec:introduction}

Sensor networks are becoming increasingly prevalent in the modern computing environment. A sensor network consists of a number of physical devices called sensor \emph{nodes}, spatially distributed within a geographic region of interest, equipped with sensing/monitoring/recording capabilities within a certain \emph{sensor  region}. Sensor networks and their data have many applications including environmental sensing, video surveillance, motion detection, and spectrum monitoring. 

\para{Motivation.}  
Our work is motivated by the problem of wireless spectrum monitoring, which is gaining momentum in the research community due to the increased availability of software defined radios (SDRs) that allow a malicious user to transmit and receive data on unauthorized spectra. Such unauthorized usage can be as simple as utilizing mobile phone networks without paying the service provider or as serious as using an SDR to distribute malware to legitimate users without their knowledge. The consequences of malware spread in this manner can be damaging to the point of causing complete network outages~\cite{KhosroshahyQiuAli2013}. Violators can even cause network disruptions by using SDRs to simply send out a jamming signal. Though cell phone jammers are illegal to purchase in the United States, users may be able to use SDRs to mimic this behavior through app-based tools. Presently, the only means for detecting spectrum violators is through reporting and manual investigation of the underlying data, which is expensive, time-consuming and ineffective. 
	
Recently, there has been a great deal of interest in leveraging crowd-sourcing of mobile phones as a kind of ad-hoc sensor network to provide robust, dynamic, and, most importantly, rapid detection of spectrum violators~\cite{AchtzehnRiihihjarviCastillo2015,DuttaChiang2016,sensor-crowd17}. 
While promising, a key challenge to the study of such sensor network data is the coverage problem. For concreteness, given a collection of nodes in a fixed, planar Euclidean domain performing some monitoring tasks, where each node has a radially symmetric (i.e.~disk-like) sensor region, we would like to understand the extent of coverage by these sensor regions. 
That is, does the union of sensor regions cover the entire domain? 
In particular, if the sensor data has location uncertainty, what is the probability of coverage within a query region? 
In the context of spectrum violator detection, for a fixed spectra, we ask the following question: does the entire domain or any queried subdomain permit a spectrum violator to go undetected, and with what probability given sensor location uncertainty?

Our work is inspired by the study of static (i.e. sensors are stationary), blanket coverage (i.e.~determine if the entire domain is covered) problem of sensor networks using topological techniques~\cite{SilvaGhrist2007}. 
When sensor locations are known deterministically, the work of de Silva and Ghrist~\cite{SilvaGhrist2007} gives a homological criterion for certifying global coverage as a novel application of ideas in topological data analysis. 
Such a problem becomes more challenging with location uncertainty in the data, which can arise in many scenarios, such as poor equipment quality, limited sensing capabilities, obstructions in the sensing environment, and periodic location estimations. 
In the case of crowd-sourcing using mobile phone networks, continually updating a sensor's location through GPS may not be practical as it increases power consumption; in addition, phone users  may not want to have their location monitored constantly or reported exactly due to privacy concerns. 
Therefore, we may only have access to intermittent or perturbed data on a phone's location, leading to a higher degree of uncertainty. 
Such uncertainty can further complicate the detection and localization of spectrum violators. 
Alternatively, we may be in a coordinate-free setting, where we are made aware of the pairwise distances between the nodes, but the locations of the sensors are unknown (e.g.~due to the preservation of privacy); in this case, we still would like to certify global coverage using topological data analysis. 

\para{Contribution.}
In this paper, we aim to provide a framework to model and visualize the network coverage problem with a simple case of location uncertainty in the setting of indecisive points~\cite{JorgensenLofflerPhillips2011}.   
Our main contribution is as follows:
\begin{itemize}\denselist
\item We model sensor network coverage with location uncertainty by quantifying the uncertainty associated with its simplicial complex representations.  
\item We introduce an interactive visualization framework that supports the exploration and manipulation of the sensor network data. 
\end{itemize}
We also discuss the educational potential of our tool to help students better understand homological criteria that originate from topological data analysis in certifying sensor network coverage. 

In addition, we have deployed our visualization tool under active development (\url{http://www.sci.utah.edu/~tsodergren/prob_net_vis_working/}) and made our code available under an open source license (\url{https://github.com/jalohse/sensor_network_topology}).  
We also provide a supplemental video to demonstrate the interactive  capabilities of our tool. 

\section{Related Work}	
\label{sec:related-work}

We review the most relevant work on two important topics, 
namely, the \emph{modeling} and \emph{visualization} of sensor network coverage, assuming either deterministic or probabilistic sensor locations. For surveys on sensor networks in general, see~\cite{AkyildizSuSankarasubramaniam2002}. 
For surveys on uncertainty visualization, see~\cite{BrodlieOsorioLopes2012, GrietheSchumann2006,PangWittenbrinkLodha1997, ThomsonHetzleraMacEachrenb2005, PotterRosenJohnson2012}. 

\para{Deterministic models of coverage using geometric methods.}
Classic geometric approaches for the coverage problem typically employ the Delaunay triangulations and require knowledge or measurement of node coordinates; see~\cite{MeguerdichianKoushanfarPotkonjak2001, LiWanFrieder2003, ZhangHou2004, FeketeKrollerPfisterer2006}. 

\para{Deterministic models of coverage using topological  methods.}
In their seminal work on sensor networks, de Silva and Ghrist\cite{SilvaGhrist2006, SilvaGhrist2007,SilvaGhrist2007b} consider the determination of coverage with minimal sensing capabilities without coordinate information. 
They demonstrate that, given a minimal set of assumptions, one can compute coverage over a compact domain through the use of simplicial complexes and persistent homology~\cite{EdelsbrunnerHarer2008}. 
Their model, while based on unknown node location, nevertheless assumes those locations are deterministic. 
Cavanna et al.~\cite{CavannaGardnerSheehy2017} recently generalized the assumptions on the boundaries to make the results applicable to general domains. 
Gamble et al.~\cite{GambleChintakuntaKrim2012} extend this concept further to consider a time-varying network. They utilize zigzag persistent homology~\cite{CarlssonSilvaMorozov2009} (a variation of persistent homology) to identify holes in the coverage area. Adams and Carlsson~\cite{AdamsCarlsson2015} use a similar approach to determine if evasion paths exist within a time-varying sensor network, that is, if a moving intruder can avoid detection in a time-varying setting.

\para{Probabilistic models of coverage.}
There is currently a limited, but growing body of work dealing with coverage issues in probabilistic sensor networks. Bhattacharya et al.~\cite{BhattacharyaGhristKumar2015} apply persistent homology to dynamic (i.e. moving) sensors in an uncertain environment for planning trajectories, where each location in the domain is assigned a probability of occupancy. 
Assuming prior knowledge on the geometry of the domain and a uniform random distribution of nodes, one can infer probability of coverage under a certain node density (e.g.~\cite{Koskinen2004, LiuTowsley2004}). 

Instead, in this paper, we draw inspirations from the \emph{indecisive} model for uncertain data~\cite{JorgensenLofflerPhillips2011}. 
In this model, each sensor node has an independent and typically distinct probability distribution of its possible locations.  These distributions could be continuous, but in this paper we focus on the case when they are discrete, described by a finite number of possible locations, for instance modeling various recent GPS readings.  We restrict that all possible locations, called \emph{indecisive points} are within a fixed shape of bounded size.  

\para{Visualizing sensor networks.}
There is a number of packages for visualizing sensor networks, see~\cite{ParbatDwivediVyas2010} for a survey. SpyGlass~\cite{BuschmannPfistererFischer2005} is a primarily communication-driven tool designed to actively monitor network health and communication links. 
Other similar interfaces include MOTE-VIEW~\cite{Turon2005},  Octopus~\cite{JurdakRuzzelliBarbirato2011} and TinyViz (for TOSSIM~\cite{LevisLeeWelsh2003}, a sensor network simulator). 
The limitation of all of these tools is that they produce mainly static displays of existing network layouts and, therefore, do not support network planning, especially in uncertain environments.
	
\para{Visualizing networks with uncertainty.}	
In the field of graph drawing, Wang et al.~\cite{WangShenArchambault2016} discuss the difficulties in dealing with ambiguities in graph layouts, and propose to identify misleading areas (created from edge bundling) to enable designers to make informed graph layout choices. 
Schulz et al.~\cite{SchulzNocajGoertler2017} combine edge bundling, splatting and a convex hull representation  to represent probability distributions of possible node locations. 

Both of these works rely heavily on enriching features of force-directed layouts. On the contrary, our visualization design focuses on quantifying uncertainty associated with sensor network coverage using simplicial complexes. 

\section{Technical Background}
\label{sec:background}

We review the topological concepts most relevant to the study of sensor network coverage; in particular, \v{C}ech complexes, Vietoris-Rips (a.k.a. Rips) complexes and homology. We define concepts in the setting of $\Rspace^2$, while most definitions generalize to high dimensions.  

We focus on an idealized sensor network model introduced in~\cite{SilvaGhrist2006}. 
Given a collection of sensor nodes in a two-dimensional domain $\Dcal \subset \Rspace^2$, we study the coverage properties of the union of (disk-like) sensor regions centered at each node. 
In the original setting~\cite{SilvaGhrist2006}, each node is coordinate-free and assumes no localization or orientation capabilities; that is, the verification of coverage is constrained by using only communication connectivity information among the nodes. 
In the context of our paper, we de-emphasize the coordinate-free aspects of the sensor network as our primary objective is to introduce  an interactive visualization framework to better understand topological approaches in studying network coverage problem with uncertainty. 

\para{Abstract simplicial complex.} 
Given a set of points in $\Rspace^2$, the smallest convex hull which contains $k+1$ vertices  $\{v_0, .., v_k\}$ is called a \emph{$k$-simplex}~\cite{Hatcher2002}. 
$0$-, $1$- and $2$-simplices correspond to vertices, edges and triangles. 
A \emph{face} is a sub-simplex of the $k$-simplex with vertex set as nonempty subset of $\{v_0, .., v_k\}$. 
A \emph{simplicial complex} $K$ is a collection of simplices such that every face of a simplex of $K$ is in $K$, and the intersection of any two simplices of $K$ is a face of each of them~\cite{Munkres1984}. 
An \emph{abstract simplicial complex} is a collection $S$ of finite nonempty sets, such that if $A$ is an element in $S$, so is every nonempty subset of $A$~\cite{Munkres1984}. 
Any abstract simplicial complex on a (finite) set of points has a geometric realization in some $\Rspace^d$.

\para{Homology.}
Given a simplicial complex $K$, roughly speaking, the $0$-, $1$- and $2$-dimensional homology of $K$, denoted as $\Hgroup_0(K)$, $\Hgroup_1(K)$ and $\Hgroup_2(K)$ respectively, correspond to the components, tunnels and voids of $K$.  
A multi-scale notion of homology is referred to as the \emph{persistent homology}~\cite{EdelsbrunnerHarer2008}, which deals with homological properties of nested families of topological spaces. 

\para{\v{C}ech complexes.}
Given a collection $\Ucal = \{U_i\}_{i=1}^{n}$ of (disk-like) sensor regions of radius $r$ centered at each sensor node ($\Ucal$ is referred to as a \emph{cover}), the \v{C}ech complex of $\Ucal$, $\Cech(\Ucal)$, is the abstract simplicial complex whose $k$-simplices correspond to nonempty intersections of $k+1$ distinct elements of $\Ucal$~\cite{SilvaGhrist2006}.  

According to the \v{C}ech Theorem~\cite{Hatcher2002}, if the cover $\Ucal$ is \emph{good} (that is, the cover sets and all finite nonempty intersections of cover sets are contractible), then the \v{C}ech complex captures the topology of the cover. In other words, the coverage properties of the cover can be captured by the properties of the \v{C}ech complex. However, computing the \v{C}ech complex is highly nontrivial, as it requires computing higher-order set intersections;  therefore we make use of a related concept, referred to as the Rips complex. 

\para{Rips complexes.}
Given a set of points in $\Rspace^2$ and a fixed radius $r > 0$, the  Rips complex, $\Rips_r$, is the abstract simplicial complex whose $k$-simplicies correspond to $k+1$-tuples of points which are pairwise within distance $r$ of each other.  The distance $r$ is therefore considered as a scale parameter.  

\para{Certifying coverage with \v{C}ech and Rips complexes.}
de Silva and Ghrist~\cite{SilvaGhrist2006} certify sensor network coverage by considering the $2$-dimensional relative homology involving Rips complexes formed by sensor nodes and Rips complexes formed by nodes on the boundary (see~\cite{SilvaGhrist2006} for technical details). 
They further extend the coverage criterion to domains of arbitrary dimensions using persistent homology~\cite{SilvaGhrist2007}.  
In a nutshell, based on our understanding of the characteristics associated with Rips and \v{C}ech complexes, we could verify coverage either from direct computation of \v{C}ech complexes (may not be practical in high dimensions or with a large number of sensors), or from structural inference based on mappings between  various Rips complexes (as approximations of \v{C}ech complexes) across different scales.  

\section{Problem Setup}

We consider a simple model of location uncertainty referred to as the \emph{indecisive data} model following the indecisive points introduced in~\cite{JorgensenLofflerPhillips2011}. 
Suppose we have a sensor network with $n$ nodes in a fixed, planar, Euclidean domain; and each node has a disk-like sensor region with a \emph{coverage radius} of $r_c$. 
Each node is associated with a finite set of $k$ possible locations (with equal probability $1/k$, independent of other nodes).
Additional, these possible locations are constrained within a disk of radius $\epsilon$ ($\leq r_c$), referred to as the \emph{disk of uncertainty}. 
The center of the disk of uncertainty is referred to as the \emph{anchor location}. 

\begin{figure}[h!]
\centering
\includegraphics[width=0.7\linewidth]{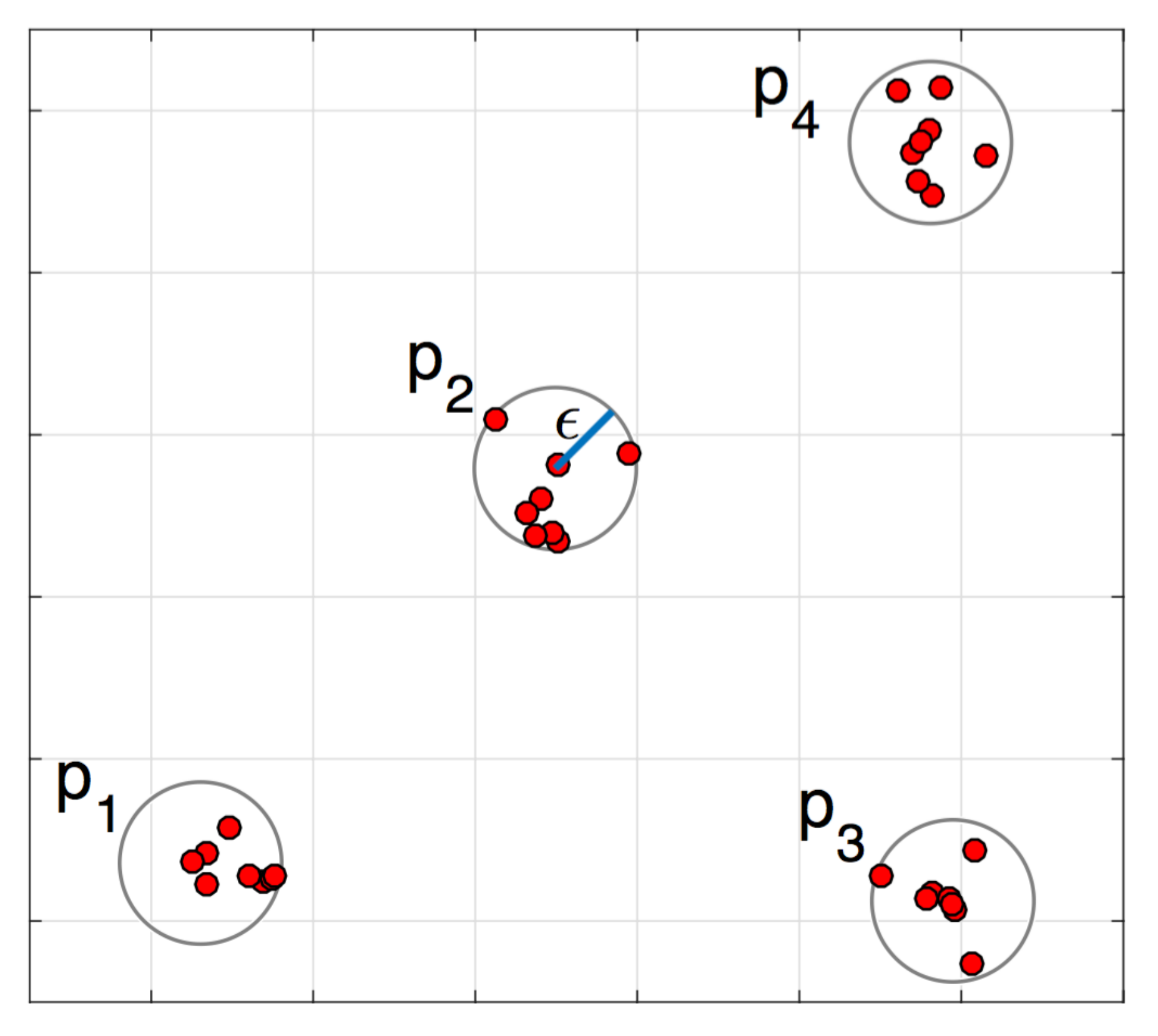}
\vspace{-2mm}
\caption{An example of a sensor network with uncertain sensor  locations under the indecisive data model, for $n = 4$ and $k = 8$. Potential sensor locations (in red) for each sensor is confined within a disk of uncertainty with radius $\epsilon$.}
\label{fig:indecisive}
\end{figure}

Formally, given a set of sensor nodes $P =  \{p_1,p_2,...,p_n\}$; each node $p_i$ has $k$ possible locations, $p_i \in \{p_i^1, p_i^2, ... ,p_i^k\}$.  
An \emph{instance} of the sensor network with $n$ nodes is one possible realization of the network where the location of each sensor is sampled among its $k$ possible locations; there are $k^n$ possible instances for a network with location uncertainty. 
An example of a simple network with $n=4$ and $k=8$ is shown in Figure~\ref{fig:indecisive}.

\section{Modeling Uncertainty}
\label{sec:model}

\begin{figure}[h]
\centering
\includegraphics[width=.65\linewidth]{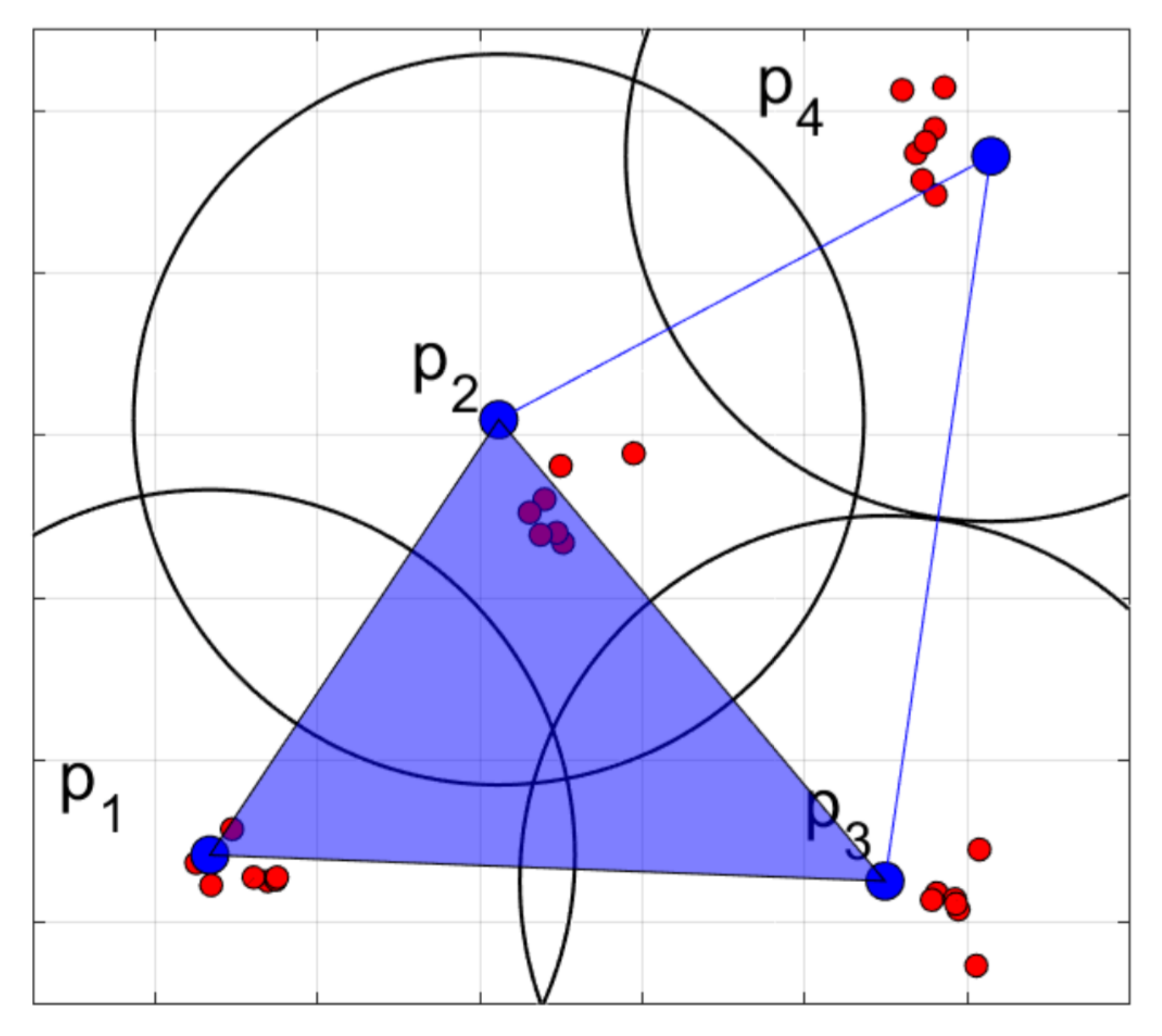}
\vspace{-2mm}
\caption{For a sensor network instance drawn from Figure~\ref{fig:indecisive} containing sensors located at $\{p_1, p_2, p_3, p_4\}$ (blue points), its corresponding \v{C}ech complex contains 1 face (i.e.~triangle $p_1p_2p_3$) and $5$ edges ($p_1p_2, p_1p_3, p_2p_3, p_2p_4, p_3p_4$). The union of disks centered at $p_2$, $p_3$ and $p_4$ does not provide local coverage, therefore no face  exists among these three points.}
\label{fig:cech}
\end{figure}

As described in Section~\ref{sec:background}, a convenient way to represent sensor network coverage comes from the notion of 
\v{C}ech complex. 
Given a sensor network with a set of nodes with coverage radius $r_c$, we can first construct its corresponding \v{C}ech complex, which captures the complete network coverage information.
That is, an edge exists between two nodes $p_i$ and $p_j$ if their respective sensor region overlaps, that is, if $d(p_i, p_j) \leq 2r_c$.
A triangle (referred to as a \emph{face}) exists among three nodes if their corresponding sensor regions share a common intersection; in other words, there exists a minimum enclosing ball of radius $\leq r_c$ that encloses all three nodes. 
In the case of a Rips complex, a face (triangle) exists among three nodes if their pairwise sensor regions overlap. For example, in  Figure~\ref{fig:cech}, the face $p_2p_3p_4$ exists for the Rips but not for the \v{C}ech complex.  
We quantify uncertainty in coverage by assigning probability measures to edges and faces in the \v{C}ech and Rips complex, respectively. 

\para{Edge probabilities for \v{C}ech complexes.}
The probability of an edge appearing between any pair of nodes $(p_i, p_j)$ across all instances can be computed by the ratio between the number of actual edge appearances and the number of possible ones.  
Under the indecisive data model, for each of the $\binom{n}{2}$ possible node pairs, there are $k^2$ possible combinations. 

\begin{figure}[h]
\includegraphics[width=\linewidth]{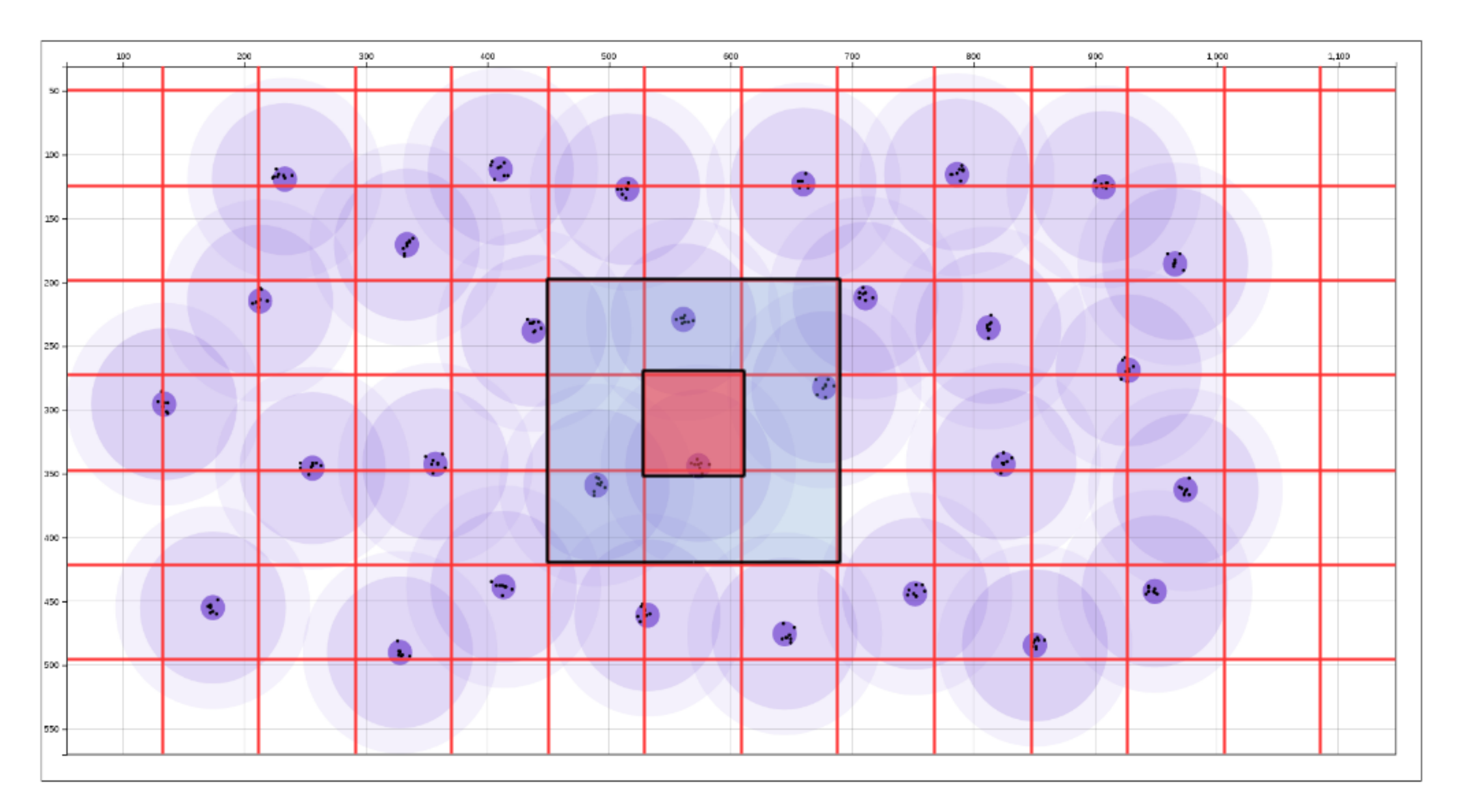}
\caption{An illustration of domain decomposition. Each block has a width equal to the maximum possible coverage diameter. One only needs to consider node pairs within the same and adjacent blocks. For example, we only need to compare nodes in the red block with nodes in the red and blue blocks.}
\label{fig:kd}
\end{figure}

We can improve the computational efficiency based on domain decomposition (similar to the idea behind the KD tree). 
This is illustrated in Figure~\ref{fig:kd}. 
We decompose the domain into square blocks of width $2(r_c+\epsilon)$ and only consider node pairs within adjacent blocks. 

\begin{figure}[h!]
\centering
\includegraphics[width=\linewidth]{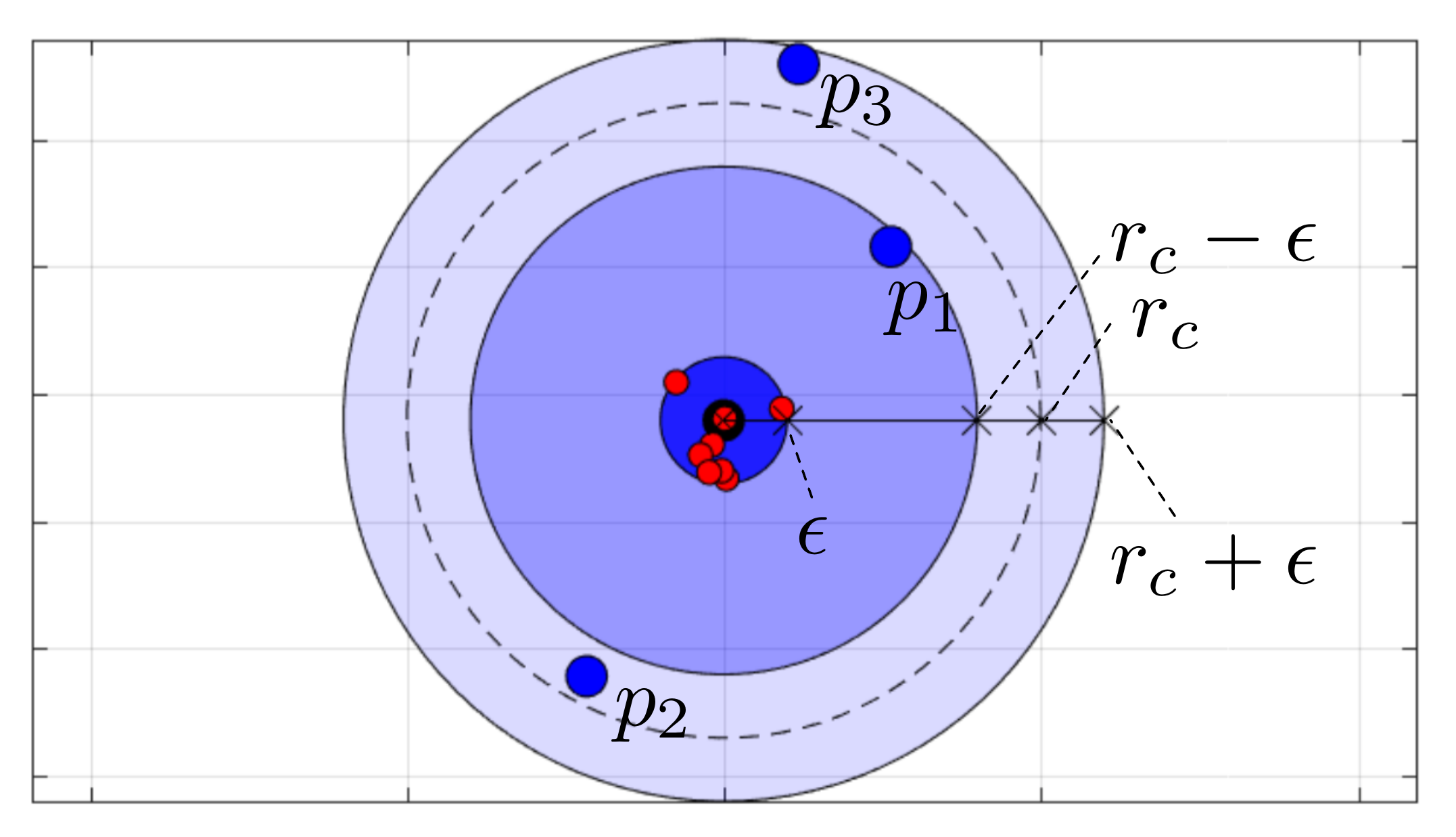}
\vspace{-2mm}
\caption{An illustration of the annulus of uncertainty (lighter shaded outer ring) surrounding a set of indecisive points (red) for one sensor node.  For any point (e.g. $p_1$) within a radius of $r_c-\varepsilon$ of the anchor point, the coverage is guaranteed, since it must be within $r_c$ of all indecisive points. For points in the annulus, shown as the lightest shade of blue (e.g., $p_2$ and $p_3$), one should consider the distance to all indecisive node locations to compute a probability of coverage. For this example, $p_2$ is covered with probability $7/8$ since $7$ of $8$ indecisive points are within distance $r_c$, while $p_3$ is covered with probability $0$ since no indecisive points are within a distance of $r_c$.}
\label{fig:annulus}
\end{figure}

Combining with domain decomposition, we could further reduce the number of node pairs under consideration by looking at nodes whose distance to an anchor point fall within so-called \emph{annulus of uncertainty}, defined between two disks of radius $r_c - \epsilon$ and $r_c + \epsilon$, see Figure~\ref{fig:annulus}. 
If a node falls inside the inner ring of this annulus, it means that all possible node pairs (with respect to the anchor point) will be connected and the edge probability is $1$. Conversely, if a node falls outside the outer ring of this annulus, it means that none of the possible node pairs will be connected and the edge probability is $0$. 
For a node that falls in between, on should consider all possible node pair combinations to compute a probability.

\para{Edge and face probabilities for Rips complexes.}
Recall that the Rips complex could be used as an approximation to the \v{C}ech complex in terms of certifying coverage. 
Given a set of nodes in a sensor network, we also model uncertainty associated with edges and faces in its corresponding Rips complex. 
Since the \v{C}ech and Rips complex share the same set of edges (therefore the same edge probabilities), we only need to infer face probabilities within the Rips complex from the edge probabilities. 

For the three boundary edges of a given potential face $p_ip_jp_k$: 
if all edge probabilities are $1$, then so is the face probability; 
if any edge probability is $0$, then so is the face probability; 
if at least two of the edge probabilities are $1$, then the face probability equals the probability of the third edge; 
otherwise, we have to iterate through all possible edges among the three nodes, where the face probability is the number of actual face appearances divided by all possible instances. 

\para{Face probabilities for \v{C}ech complexes.}
To compute face probabilities for \v{C}ech complexes,  we rely on the iterative procedure similar to the one used for Rips complexes; that is, the face probability among three fixed sensor nodes is the number of actual face appearances in the \v{C}ech complex divided by the number of possible faces.

\section{Visualization Design}
\label{sec:vis}

Our visual interface contains four panels (Figure 1): control panel (a), viewing panel (b), interaction panel (c) and color panel (d). 

\begin{figure*}[h!]
\centering
\includegraphics[width=.95\linewidth]{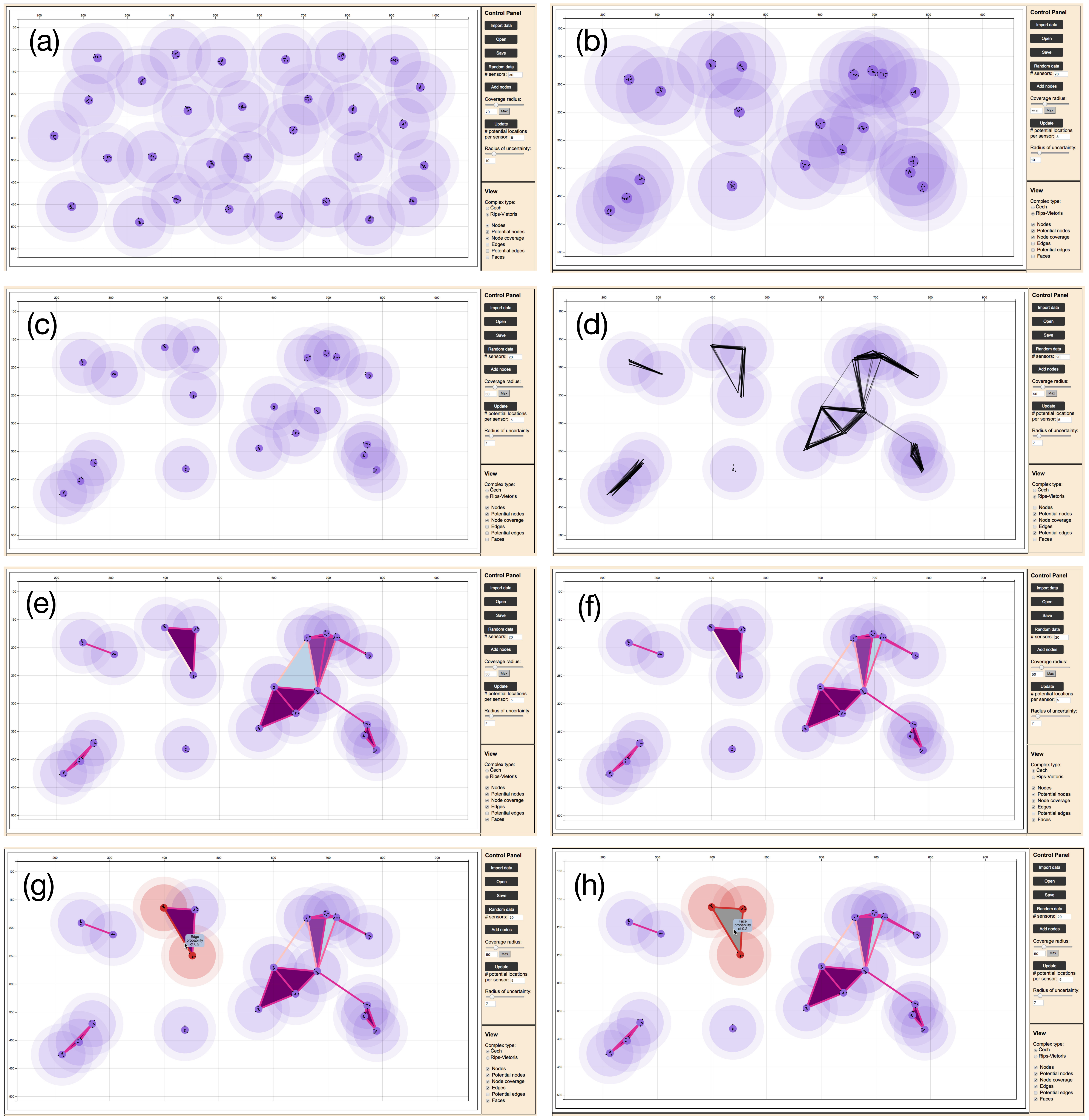}
\vspace{-2mm}
\caption{Various visualization capabilities with our tool. (a) A randomly generated sensor network with default configurations. (b) A user-specified data set loaded with CSV format. (c) Adjusting radius of coverage from (b) to the current configuration. (d) Displaying all possible edges.  (e) Rips complex for the configuration in (c). (f) \v{C}ech complex for the configuration in (c). (g)-(h) Hovering over an edge and a face respectively will highlight relevant elements in the complex and display the corresponding probabilities.}
\label{fig:usecase}
\end{figure*}

\para{Control panel.} 
The control panel contains adjustable configuration parameters, 
including the number of sensors $n$ (default $n = 30$), the number of potential locations per sensor $k$ (default $k = 8$), the coverage radius $r_c$ and the radius of uncertainty $\epsilon$ (default $\epsilon = 10$, max $50$). 

Users can generate $n$ uniformly random distributed sensors in the domain of the interaction panel (via the \emph{random data} button, Figure~\ref{fig:usecase}(a)).
Users have the ability to load existing datasets (e.g.~generated by simulations) in CSV format containing x and y coordinates of sensor nodes (via the \emph{import data} button, Figure~\ref{fig:usecase}(b)), or in JSON format encoding both coordinates and configuration parameters (via the \emph{open} button). 
The existing network configuration within the interaction panel can be saved in JSON format as well (default as data.json via the \emph{save} button).

The coverage radius $r_c$ and the radius of uncertainty $\epsilon$ can be adjusted using a slider or specified in a text box (Figure~\ref{fig:usecase}(c)).  
Unless specified in a pre-loaded JSON file, the maximum value for $r_c$ is initialized as half of the largest distance between sensors along the x-axis. This maximum value can also be updated (via the \emph{max} button). 
Once the number of potential locations per sensor $k$ is specified, the configuration is updated by generating $k$ random locations within the disk of uncertainty via the \emph{update} button.

\para{Viewing panel.}
Via the viewing panel, users can switch between \v{C}ech or Rips complex of the current network configuration (default to Rips complex), see Figure~\ref{fig:usecase}(e)-(f) for Rips and \v{C}ech complex respectively. 

Various layers of information can be super-imposed onto each other within the interactive panel via various checkboxes. 
The \emph{nodes} box determines whether a sensor is shown with its radius of uncertainty. Checking the \emph{potential nodes} box displays the randomly generated potential locations for each node. The \emph{node coverage} box enables the visualization of coverage region per node centered at the anchor point (in shades of purple); the inner circle of such a coverage area designates the node coverage radius $r_c$, while the outer circle represents the coverage radius plus the radius of uncertainty $r_c+\epsilon$. 
Edges and faces in the simplicial complex are shown via the \emph{edges} and \emph{faces} box respectively. 
Users also have the ability to view all possible edges via the \emph{potential edges} box, that is, all edges that exist between potential locations within the coverage radii (Figure~\ref{fig:usecase}(d)). 

\para{Color panel.}
The color panel is used to visualize edge and face probabilities and is rather self-explanatory. Seven and five color maps with color discretization are available for visualizing face and edge probabilities respectively. 

\para{Interaction panel.}
Within the interaction panel, users have the ability to move, delete, and add nodes. Moving an existing node can be accomplished by mouse clicking, dragging and releasing.  
Adding a node is done via the \emph{add node} button in the control panel. When a node is selected via mouse clicking, it can be deleted with the \emph{delete} key.
When a node is selected, the nodes and its adjacent edges and faces are highlighted in shades of red; while its neighboring nodes are highlighted in blue. Deselecting a node is done with the \emph{esc} key. 

\begin{figure}[h!]
\centering
\includegraphics[width=.98\linewidth]{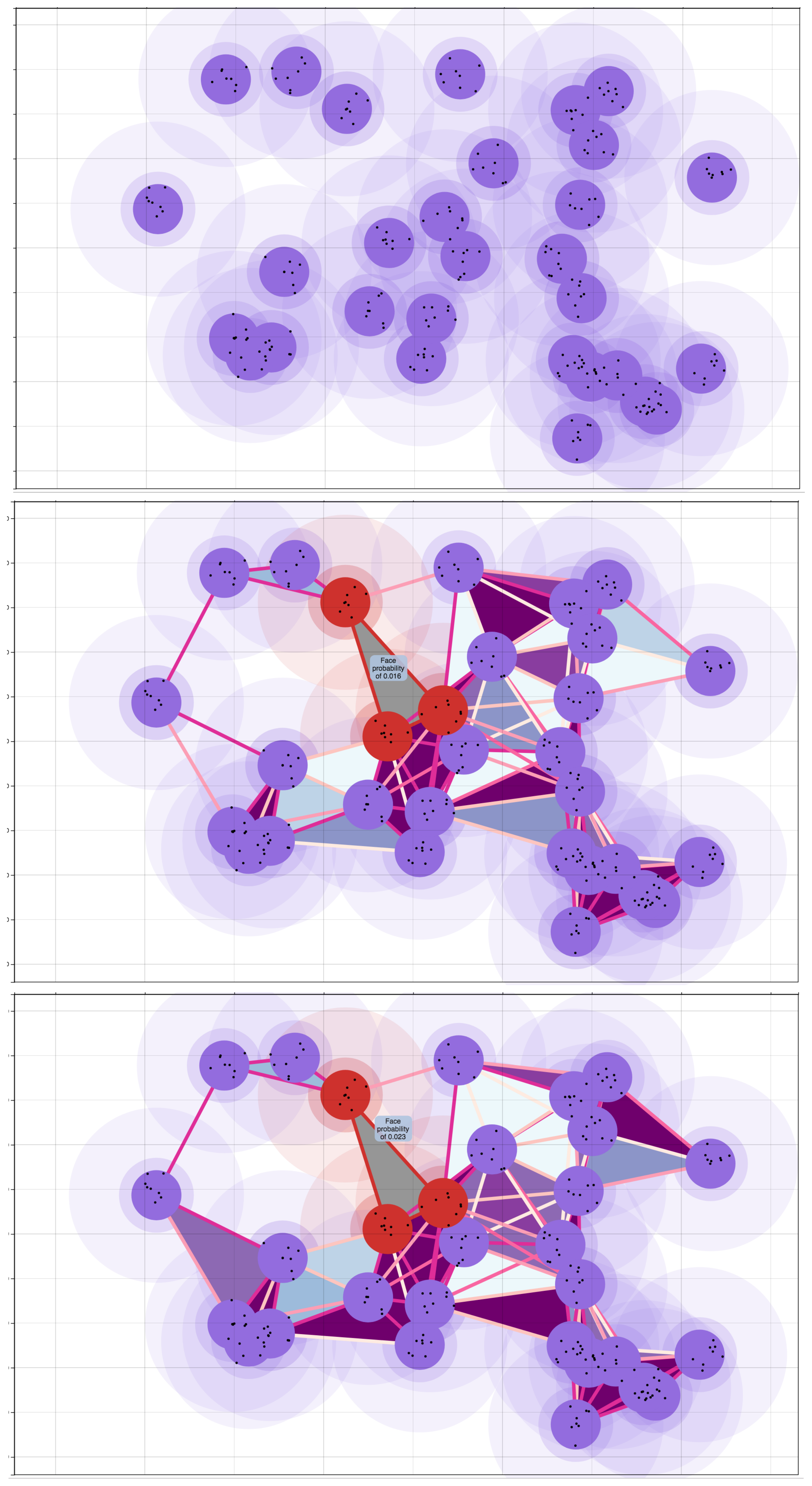}
\vspace{-2mm}
\caption{Highlighting the differences among face probabilities for the \v{C}ech (middle, face probability: $0.016$) and Rips complex (bottom, face probability: $0.023$) respectively.}
\label{fig:facehighlight}
\end{figure}

\begin{figure}[h!]
\centering
\includegraphics[width=.6\linewidth]{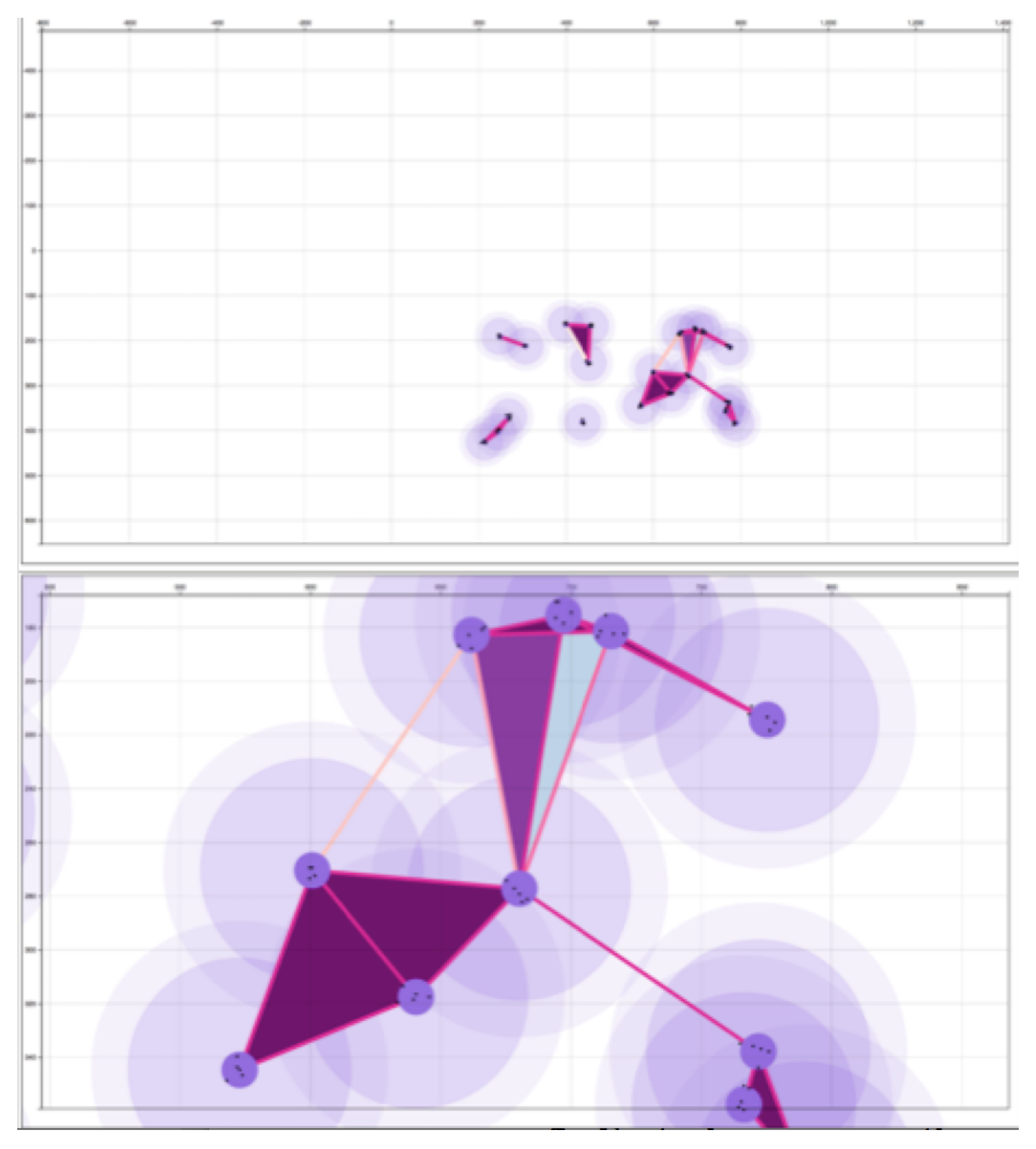}
\vspace{-2mm}
\caption{A zoomed out (top) configuration and a zoomed in (bottom)  configuration.}
\label{fig:zoom}
\end{figure}

When a user's mouse hovers over a particular simplex (an edge or a face), the simplex and its vertices are highlighted in red. 
Hovering over a face or an edge also shows a small tooltip that displays its corresponding face or edge probability, see Figure~\ref{fig:usecase}(g)-(h) for different visualization capabilities. 
We further use Figure~\ref{fig:facehighlight} to demonstrate the differences in face probabilities based on either \v{C}ech or Rips complex for the same sensor network. 
Finally, the interaction panel supports zooming operations, as illustrated in Figure~\ref{fig:zoom}. 

\begin{figure}[ht!]
\centering
\includegraphics[width=.98\linewidth]{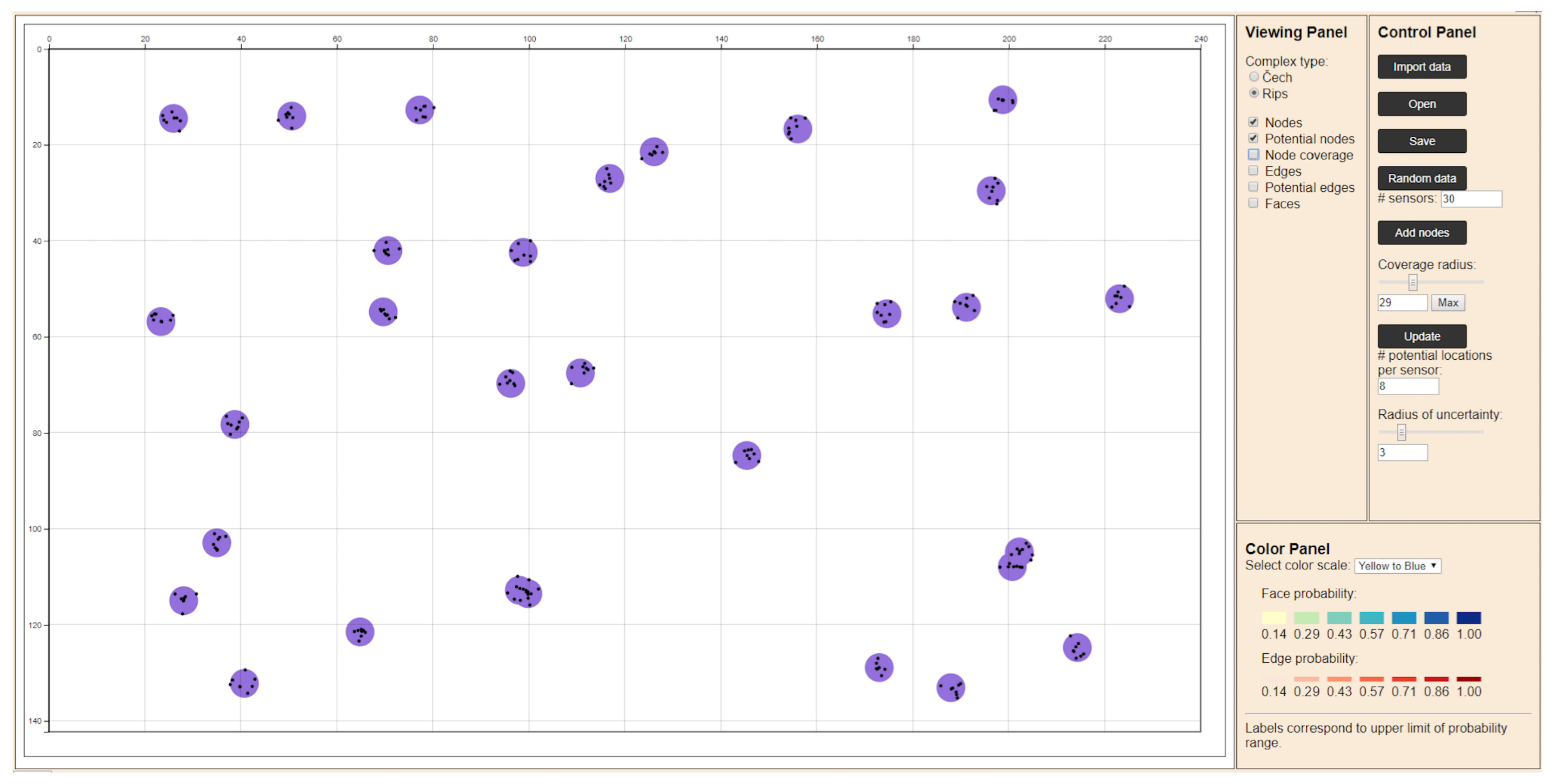}
\vspace{-2mm}
\caption{An initial network configuration for the spectra violator detection.}
\label{fig:campus}
\end{figure}

\section{An Example Workflow} 
We demonstrate an example workflow with our tool using a pre-defined sensor network configuration (e.g.~$n=30$, $k=8$, $r_c=29$, $\epsilon = 3$).
We discuss several instances of analyzing a given sensor network in order to illustrate the usability of our tool. 
Suppose we would like to utilize a mobile phone network via crowd-sourcing to detect malicious users on campus that transmit and receive data on unauthorized spectra. The initial network configuration highlighting sensor (i.e. mobile phone) locations with uncertainty is illustrated in Figure~\ref{fig:campus}.

\begin{figure*}
\centering
\includegraphics[width=.85\linewidth]{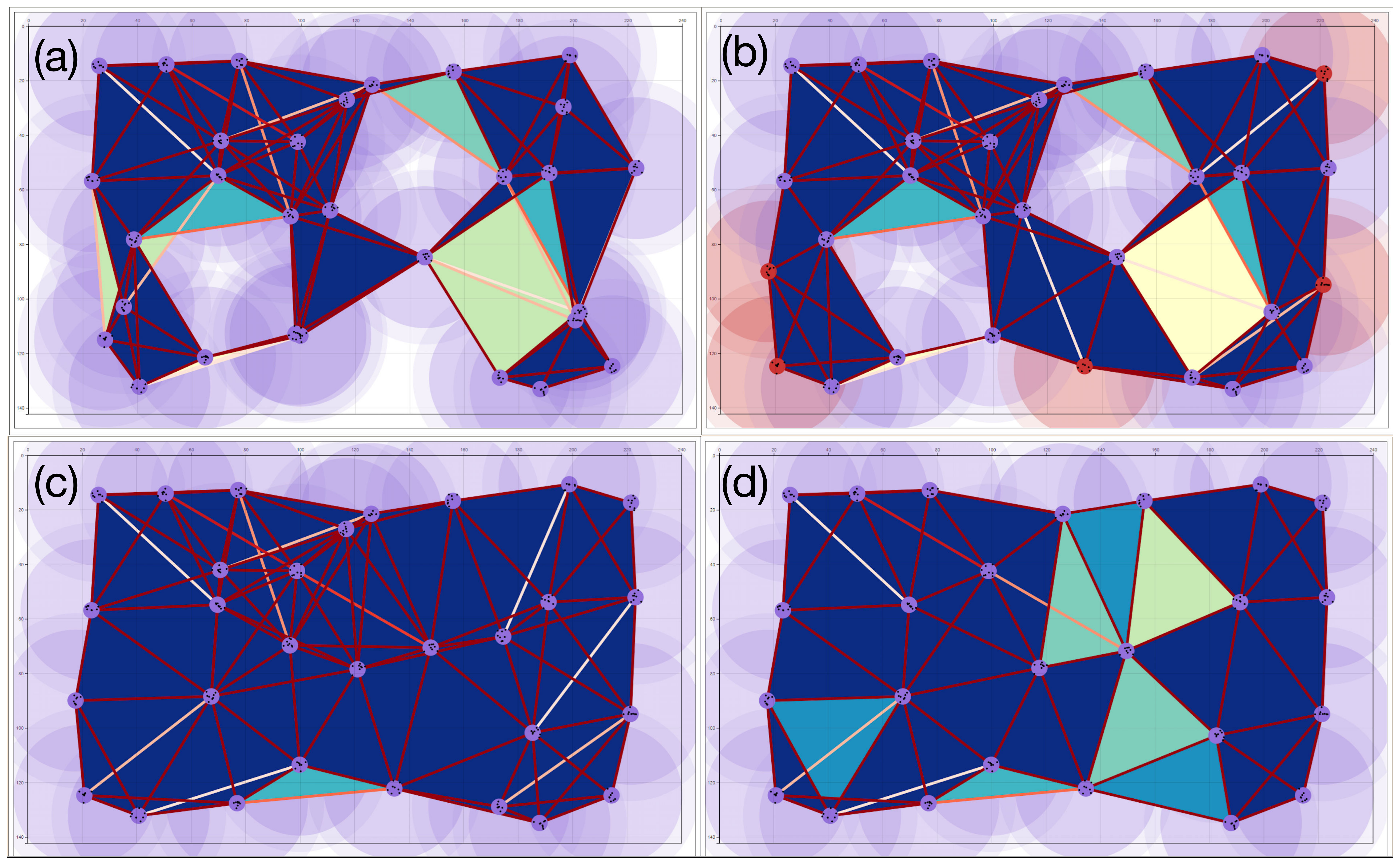}
\vspace{-2mm}
\caption{Analyze, modify and sparsify network configuration to guarantee coverage.}
\label{fig:campus-coverage}
\end{figure*}

First, with a fixed network configuration, we would like to determine whether we have coverage for the entire domain (e.g.~the campus) or for a particular subdomain (e.g.~areas surrounding the administrative building). The spectra coverage criteria can be quickly evaluated via its \v{C}ech complex representation in Figure~\ref{fig:campus-coverage}(a), where there are clearly two uncovered areas in the interior together with area along the campus boundary. 
These uncovered areas are the vulnerable regions in the domain where a malicious spectrum violator can hide.  

Second, we may wish to modify an existing sensor network configuration with minimal equipment overhead to achieve coverage. To start, we may move existing sensors around to establish a ``fence" along the campus boundary, see Figure \ref{fig:campus-coverage}(b) where we use an intuitive approach by moving nodes closest to the boundary without disturbing its interior coverage configuration. 
Then we adjust (or in some cases, add) a few interior nodes to increase coverage probability or to achieve (almost) global coverage in Figure \ref{fig:campus-coverage}(c). With such coverage, almost no spectrum violator can escape detection. 

Finally, we could \emph{sparsify} the sensor network by removing redundant sensors from dense regions without affecting the domain coverage. We could even sparsify further if we tolerate lower probability of coverage in certain subdomains, see Figure \ref{fig:campus-coverage}(d). This way we could reduce energy and equipment cost in crowd-sourcing that achieve the same effective coverage. 

\begin{figure}[h!]
\centering
\includegraphics[width=.98\linewidth]{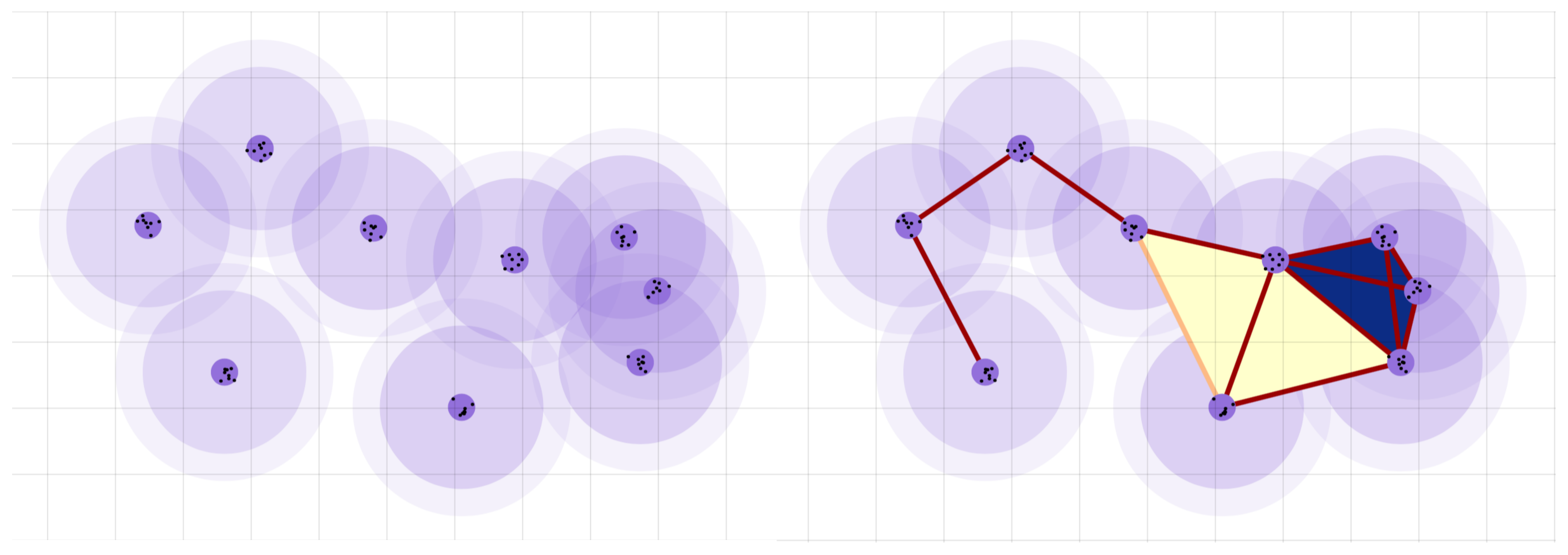}
\vspace{-2mm}
\caption{The \v{C}ech complex of the sensor regions (right) captures the homotopy type of the cover (left).}
\label{fig:homotopy}
\end{figure}

\section{Education}

Our visualization interface has the potential to be used as an educational tool to explore topological approaches in modeling the  network coverage problem. We discuss several aspects here in studying the blanket coverage problems with computational topology. 

We consider the case where sensor nodes lie within a planar Euclidean domain. The problem of \emph{blanket coverage} is concerned with the following problem~\cite{SilvaGhrist2007b}: Does the union of the sensor regions cover a given domain? 

\para{Homotopy type.} 
First, by exploring the \v{C}ech complex representation of the network configuration, one could have a good understanding of how the \v{C}ech complex of a cover by convex sets (here, disk-like sensor regions) can capture the homotopy type of the cover itself, see Figure~\ref{fig:homotopy}.

\para{Optimal factorization of the Rips complex.}
Second, by exploring relations between \v{C}ech and Rips complexes, one could study the optimal factorization of the Rips complex~(Theorem 2.5, \cite{SilvaGhrist2007}). That is, the nesting of a \v{C}ech complex between a pair of Rips complexes of varying radii. 
Formally, the \v{C}ech and the Rips complex of a set of sensors in $\Rspace^2$ gives the following chain of inclusions for $r/r' \geq \sqrt{\frac{4}{3}}$: 
\begin{equation*}
\Rips_{r'} \subset \Cech_{r} \subset \Rips_{r}.
\end{equation*}  

\para{Topological coverage criterion.} 
Third, with further development of the tool, one could potentially use our tool to explore various domain assumptions of the \emph{Topological Coverage Criterion} (TCC)~\cite{SilvaGhrist2007, CavannaGardnerSheehy2017}. 
One can explore various network configurations where homological criterion holds (or does not hold) and study the cause of failure cases via their Rips complexes. 
One could also explore sparsification schemes that simplifies a redundant cover using appropriate choice of generators~\cite{SilvaGhrist2007}. 
If our tool could be extended to include polygonal boundaries instead of just rectangular ones, one could explore further the various boundary configurations within TCC~\cite{CavannaGardnerSheehy2017}.

\section{Discussions}

We introduce an interactive framework that models and visualizes sensor networks with location uncertainty. We assign probabilistic measures to simplicial complexes that capture or approximate network coverage. Our visualization interface explores topological approaches in certifying global and local network coverage. It also enables the manipulations of parameters to better understand the robustness of coverage among various network configurations.

\para{Scalability.} We do not emphasize here the computational issues. However our current implementation suggests room for improvement in terms of scalability. The current iterative procedure in modeling and visualizing uncertainty could handle roughly hundreds of nodes. 
Realistically-sized networks, for example, wireless sensor networks, may consist of thousands, tens of thousands or millions of devices. 
For instance, over 90\% of US adults have cell phones, and if most of them (through carrier auto-enrollment) are in a network that could be used for spectra violator detection, that is well over 100 million devices.   
Modeling and visualizing any larger networks require drastically different approaches, including potentially distributed computation, node sparsification and structural summarization. 

\para{Extensions.} An immediate extension to the current platform is to better represent, compute and visualize complexes capturing the various boundary conditions. The long term objective is to employ the probabilistic notions on simplicial complexes for advanced, topology-based analysis of sensor networks in the time-varying settings. 

\section*{Acknowledgements}
The authors wish to thank Dan Maljovec for providing some source code during the initial phase of this project. 
This work was supported in part by NSF IIS-1513616, CCF-1350888, IIS-1251019, ACI-1443046, CNS-1514520, and CNS-1564287.

\bibliographystyle{abbrv}
\bibliography{sn.bib}

\end{document}